\documentclass[aip,reprint]{revtex4-1}

\usepackage{amssymb}
\usepackage{amsmath}
\usepackage{mathtools}
\usepackage{graphicx}
\usepackage{multirow}
\usepackage{longtable}
\usepackage{threeparttable}
\usepackage[colorlinks,linkcolor=blue,anchorcolor=black,citecolor=blue,urlcolor=black]{hyperref}
\graphicspath{{./fig/}}

\begin{document}


\title[Symmetry-Preserved Artificial Intelligent]{A Symmetry-orientated Divide-and-Conquer Method for Crystal Structure Prediction}



\author{Xuecheng Shao}
\author{Jian Lv}
\author{Peng Liu}
\author{Sen Shao}
\author{Pengyue Gao}
\author{Hanyu Liu}
\affiliation{International center of computational method and software, College of Physics, Jilin University, Changchun 130012, China}
\author{Yanchao Wang}
\email{wyc@calypso.cn}
\affiliation{International center of computational method and software, College of Physics, Jilin University, Changchun 130012, China}
\affiliation{State key laboratory of superhard materials, College of Physics, Jilin University, Changchun 130012, China}
\author{Yanming Ma}
\email{mym@jlu.edu.cn}
\affiliation{International center of computational method and software, College of Physics, Jilin University, Changchun 130012, China}
\affiliation{State key laboratory of superhard materials, College of Physics, Jilin University, Changchun 130012, China}
\affiliation{International Center of Future Science, Jilin University, Changchun 130012, China}

\date{\today}

\begin{abstract}
Crystal structure prediction has been a subject of topical interest, but remains a substantial challenge, especially for complex structures as it deals with the global minimization of the extremely rugged high-dimensional potential energy surface. In this manuscript, a symmetry-orientated divide-and-conquer scheme was proposed to construct a symmetry tree graph, where the entire search space is decomposed into a finite number of symmetry-dependent subspaces. An artificial intelligence-based symmetry selection strategy was subsequently devised to select the low-lying subspaces with high symmetries for global exploration and in-depth exploitation. Our approach can significantly simplify the problem of crystal structure prediction by avoiding exploration of the most complex $P1$ subspace on the entire search space and have the advantage of preserving the crystal symmetry during structure evolution, making it well suitable for predicting the complex crystal structures. The effectiveness of the method has been validated by successful prediction of the candidate structures of binary Lennard-Jones mixtures and high-pressure phase of ice, containing more than one hundred atoms in the simulation cell. The work, therefore, opens up an opportunity towards achieving the long-sought goal for crystal structure prediction of complex systems. 
\end{abstract}

\maketitle 


\section{Introduction}
Knowledge of crystal structures is essential if the properties of materials are to be understood and exploited, particularly when establishing a correspondence between materials performance and their chemical compositions. There is a high interest in crystal structure prediction (CSP), where crystal structures are precisely predicted from theory without acquiring any prior known structure knowledge, in case the only given information is the chemical compositions of materials. Thermodynamics plays a critical role in determining the structures and the likelihood of the structures formed in nature associate with their energies. CSP is targeted to identify the energetically most favourable structure that is synthesizable in experiments and whose energy is a global minimum on the potential energy surface (PES), a vast ``landscape'' in a high-dimensional space that possesses high energy barriers separating energy minima.

Since the analytic form of the PES is unfortunate unknown, a numerical solution for finding the global minimum is essential. Application of the variable-cell geometry optimization\cite{nocedal2006numerical}, which is commonly used in modern CSP methods\cite{Call2007, Pickard_2011, Wang_2012, wang2010crystal, gao2019interface}, simplifies the targeted PES from a continuous landscape into discrete energy minima. However, the number of energy minima is still in an astronomical figure (e.g., it is roughly estimated to be $10^{42}$ for a system of 100-atom Lenard-Jones cluster\cite{forman2017modeling}) and scales exponentially with the number of atoms in a structure. Mathematically, global minimization among these energy minima is a nondeterministic polynomial-time hard problem, posing a grand challenge for CSP\cite{oganov2018crystal}.

A variety of popular CSP methods (see e.g., Ref. \citenum{wang2014perspective} for details on the different methods) were recently developed and successfully applied to solve structure-related problems, leading to a number of major discoveries (e.g., the finding of pressure stabilized high-temperature superconductor LaH$_{10}$ that holds the record high $T_c$ at 260 K known thus far \cite{peng2017hydrogen, liu2017potential}). These methods were proposed based on various samplings schemes on PES, including the simulated annealing\cite{Brooks1995}, basin-hopping\cite{Wales1997}, minima hopping\cite{Amsler_2010}, metadynamics\cite{Marto_k_2003}, random sampling\cite{Pickard_2011}, genetic algorithm\cite{Woodley1999,Oganov_2006,Abraham_2006,Trimarchi2007,Lonie2011} and swarm-intelligence algorithm\cite{Call2007, Wang_2012, wang2010crystal}.
These methods use different structure searching schemes but share a common strategy: direct sampling over the entire PES\cite{oganov2019structure}. Since the search space is vast as aforementioned and the typical first-principles structure searching simulations can only explore several ten thousand structures or much less, a direct sampling faces inevitably a problem of insufficient sampling especially for a large system (e.g., structures having $>$50 atoms in the unit cell). 

Crystal structures sitting at PES constitute a vast structure (or configuration) space. A sampling on structure space is mathematically equivalent to sampling on PES. In an effort to avoid the above-mentioned insufficient problem for a direct sampling on PES associating with the spatial arrangements of atoms, we develop a symmetry-orientated divide-and-conquer scheme via the construction of a symmetry tree graph (STG) that allows a rigorous decomposition of a vast structure space into a number of symmetry dependent structure subspaces and elimination of the most complex $P1$ subspace to significantly reduce the complexity of structure space\cite{cormen2009introduction}. A symmetry-preserved artificial intelligent algorithm (SPAI) was subsequently devised to locate the suitable subspaces and further perform in-depth exploitation in the selected promising subspaces. As we will illustrate in more detail below, if the stable crystal structure has a high symmetry, high searching efficiency and success rate can be achieved for the current method.

\section{Methods}

\begin{figure*}[htp]
	\centering
	\includegraphics[width=0.8\linewidth]{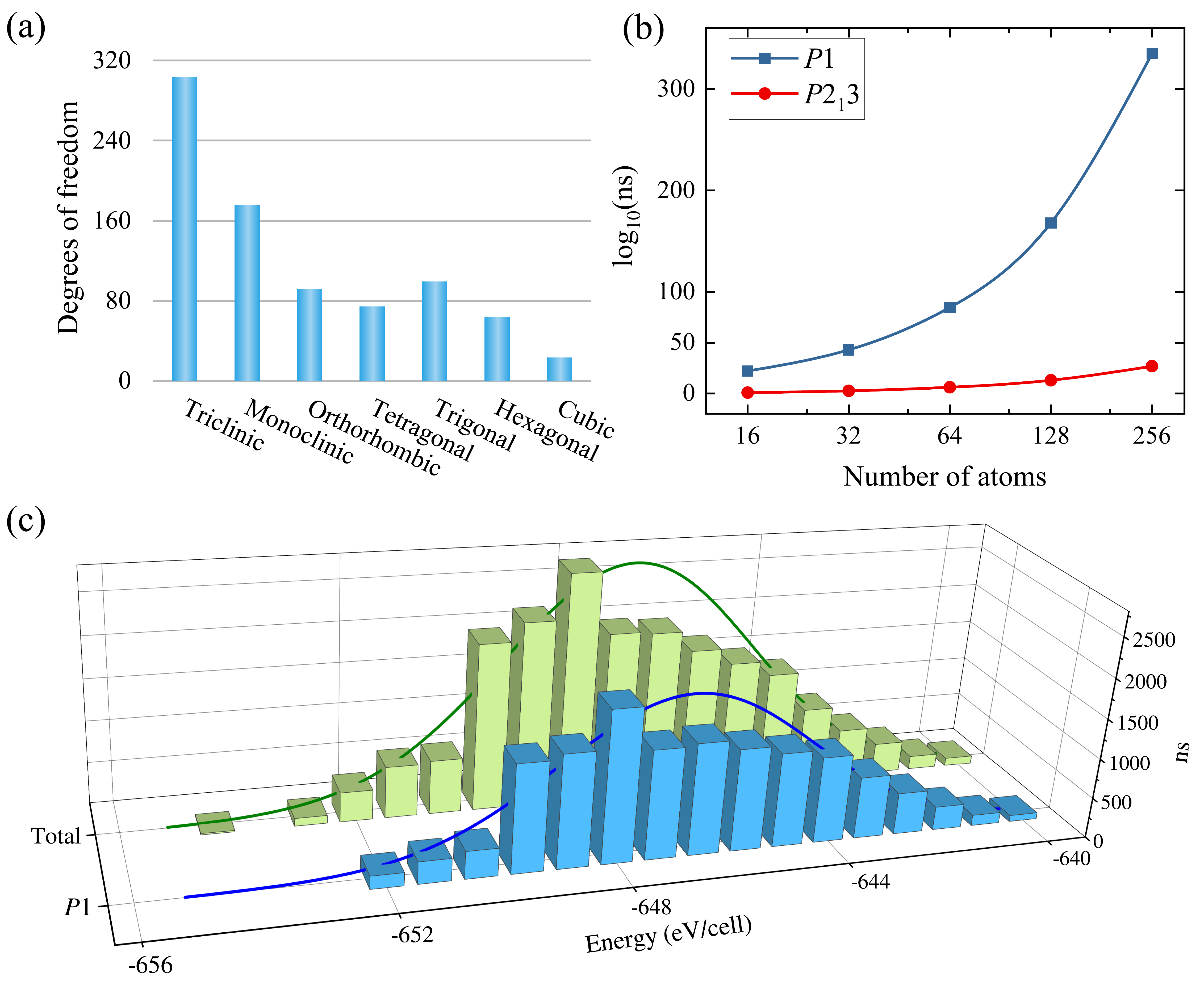}
	\caption{\label{fig:num} (Color online) (a) The maximal number of degrees of freedom to model a crystal structure containing 100 atoms for the choices of different symmetries. (b) The estimated number of structures ($ns$) in a $log_{10}$ scale sitting at energy minima versus the various system sizes for $P1$ and $P2_{1}3$ symmetries. Note that the system-specific constant of $\alpha$ is set to be 1.0 for both systems. (c) The distribution of 20,000 random structures of MgAl$_2$O$_4$ over the energy for all symmetries and $P1$.}
\end{figure*}

Description of a crystal structure containing $N$ atoms needs a maximal number of $3N+3$ degrees of freedom for structural parameters including 6 for the crystallographic unit cell and $3N-3$ for atomic positions. The actual number of degrees of freedom depends on the symmetry of a structure. As depicted in Fig. \ref{fig:num}(a), for a 100-atom structure, 303 degrees of freedom are required to model a triclinic system, while it is substantially reduced to 23 for a high symmetric cubic system. It is empirically suggested that the search space with a high degree of freedom usually has a large hyper-volume\cite{massen2007power} on PES, as further supported by the mathematical fact that there is an exponential increase of the number of energy minima $n(d)$ with the number of degrees of freedom ($d$): $n(d)={\rm e}^{\alpha d}$, where $\alpha$ is a system-specific constant\cite{oganov2018crystal}. The dramatic reduction of $d$ for a high symmetric structure results in a substantially reduced $n(d)$ as illustrated in Fig. \ref{fig:num}(b), where $n(d)$ for a low symmetric $P1$ structure is compared with that for a high symmetric $P2_{1}3$. It is seen that the number of energy minima for a 64-atom system reaches up to approximately $10^{85}$ for $P1$, whereas it is amazedly reduced to $10^{6}$ for $P2_{1}3$.

As described above, the complexity of structure space is originated from the complexity of low symmetric structures. As a result, a direct sampling over structure space inevitably enhances the visibility of low-symmetric structures, whereas high-symmetric structures are underrepresented as illustrated by a numerical experiment for 20,000 random structures of MgAl$_2$O$_4$ showing 85.85\% and 14.15\% occupancies for $P1$ and other symmetric structures in Fig. \ref{fig:num}(c), respectively. In case that the true structure is having a high symmetry, the problem we face for CSP would be much simplified as we can mainly focus on high symmetric structures. Our wish is not in contradiction with the previous statistical analysis on that the crystal structure is likely with high symmetry\cite{Wales_1998}. 

\begin{figure*}[htp]
	\centering
	\includegraphics[width=0.8\linewidth]{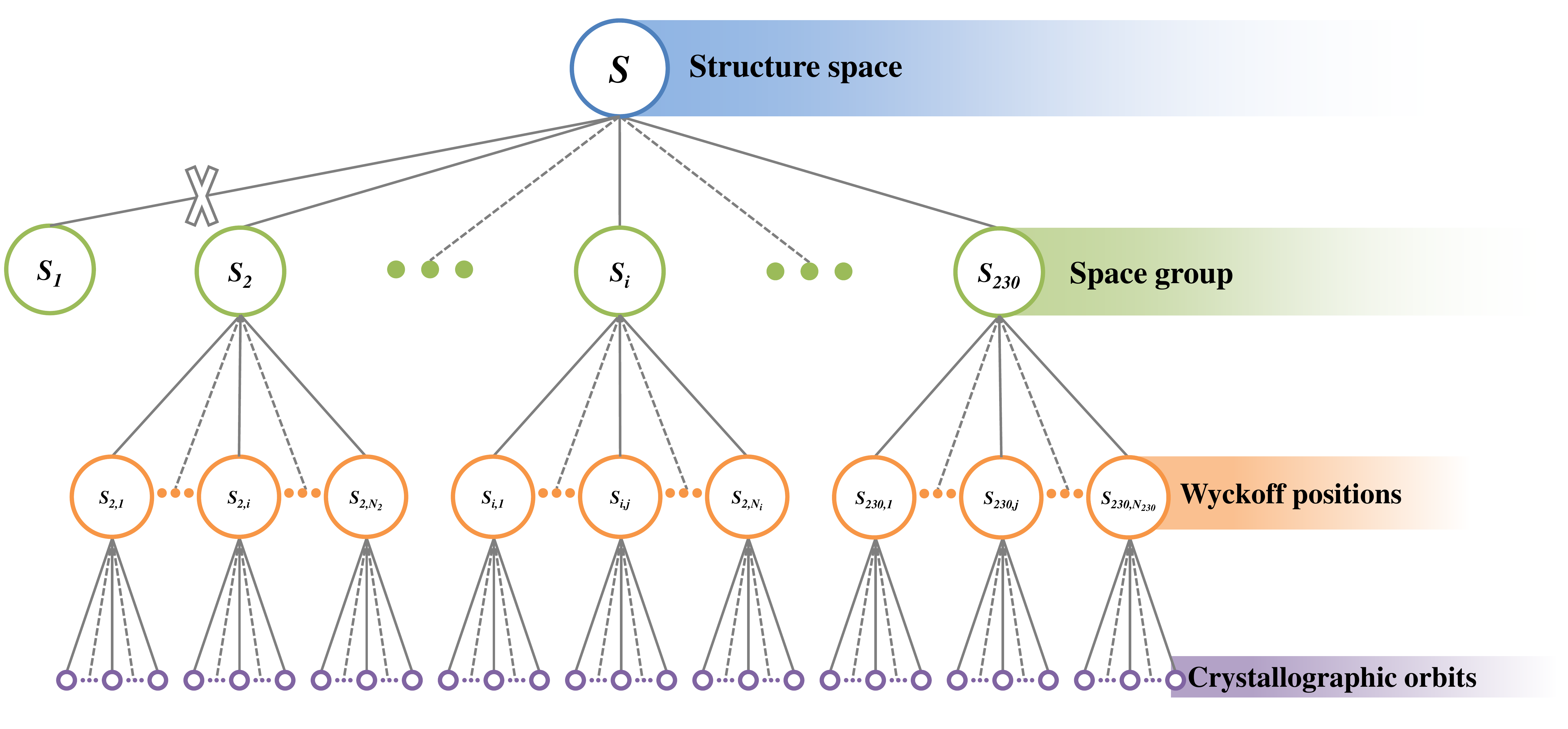}
	\caption{\label{fig:STG} (Color online) A schematic representation of STG and SPAI for the exploration of STG. Note that each node in STG is located along a one-dimensional unphysical coordinate simply for visual clarity.}
\end{figure*}

Earlier methods\cite{Wang_2012, Lyakhov_2013} that use symmetry for the generation of structures can not be used for such a purpose since the symmetry is not preserved during structure evolution, leading to enhancement of the visibility of $P1$ structures. While the predefined cell parameters are usually required to preserve symmetry for several genetic algorithms using symmetry-adapted-crossover operations. Thus, these methods can only apply to specific systems (e.g., substitutionally disordered materials)\cite{mohn2011predicting}. We here construct a three-level STG in Fig. \ref{fig:STG} to represent the structure space, in which crystal structures are grouped into a set of crystal symmetry dependent subspaces (level 1) and site-symmetry related groups (level 2) following the rules of crystallographic symmetry. In level 1, for a three-dimensional crystal, using 230 space groups as subspaces can give a rigorous description of an entire structure space (S) as described by $S=\bigcup_{i=1}^{230}S_i$, where $S_i$ denotes the $i$-th subspace with the $i$-th space group. Each subspace $S_i$ can be further subdivided into a set of site-symmetry related groups ($S_{i,j}$) in level 2 as described by $S_i=\bigcup_{j=1}^{n}S_{i,j}$. Within each $S_{i,j}$ the structures share the same combination of the Wyckoff positions. The complete list of all possible combinations ($n$) is mathematically enumerated and atomic positions in structures can be obtained by coordinate descriptions of the Wyckoff positions, i.e., crystallographic orbits. In level 3, once the crystallographic orbits are determined, the structure space is eventually decomposed into symmetry-catalogued subspaces within which the structures could reduce to a same local minimum structure after the geometry optimizations. One example of STG of MgAl$_2$O$_4$ which has 28 atoms in the cell was presented in Fig. S1.

With the STG at hand, it is now possible to develop the divide-and-conquer scheme for crystal structure prediction. There is a need of three different agents that allow for proper samplings on the corresponding three different levels in the STG. We name them as scout, onlooker, and employee agents, which are borrowed from the artificial bee colony algorithm\cite{zhang2015abcluster, zhang2016global, yanez2018automaton}. Our method is a population-based evolutionary scheme, in which the initial structures in the first population are generated randomly with the symmetry constraints. Note that the candidate structures with $P1$ space group are excluded to reduce the complexity of structure space. All structures are optimized and ranked in order of their energies as high, middle, and low energy structures that are then assigned as scouts, onlookers, and employees, respectively. Structures in the next population are generated with the aid of all three agents of scout, onlooker, and employee.

The agent scout is responsible for the exploration of 229 subspaces not including $P1$ in level 1 of STG. Scouts are first discarded and then re-generated by randomly choosing space groups to avoid any personal bias on the generation of structure. At the same time, a strict control to avoid the repetition of the same space group has been imposed until all 229 space groups have been examined. These constraints ensure the samplers walk over less-explored space groups for better coverage of the entire structure space. The agent onlooker is responsible for the exploration of site-symmetry related groups in level 2 of STG. With the information of space group unaltered, onlookers randomly choose a different combination of the Wyckoff positions allowed following the probability $p_i$.
\begin{equation}
	p_{i}=\frac{fit_{i}}{\sum_{i=1}^{SN}fit_{i}},
\end{equation}
where $SN$ denotes the number of onlookers, and $fit_{i}$ is evaluated by its energy ($E_{i}$):

\begin{equation}
	fit_{i}=
	\begin{dcases}
		\frac{1}{1+E_{i}} & \text{if } E_{i}\geq 0; \\
		1+\lvert E_{i} \rvert & \text{if } E_{i}< 0
	\end{dcases}
\end{equation}

The use of probability control ensures that onlookers with lower energies have a higher probability to be selected for the generation of structures in the next population. The agent employee is responsible for the exploration of crystallographic orbits in level 3 of STG. With the information of the space group and site-symmetry group unaltered, employees randomly choose different atomic coordinates of the Wyckoff positions to generate structures in the next population. Our structure searching scheme is controlled in a self-organized manner and the roles of three agents can dynamically change depending on the order of their energies. When an employee cannot be further improved within certain predetermined cycles, it automatically becomes a scout, whereas a scout with lower energy can change its role as an employee or onlooker. The structural variations of onlooker and employee act, from the point of view of the entire population, as feedback, amplify the promising structure space by sharing their crystallographic information, and ensure the structures in the population evolving positively by performing more attempts nearby the low-lying structure space.

\begin{figure}[htbp]
	\centering
	\includegraphics[width=0.85\linewidth]{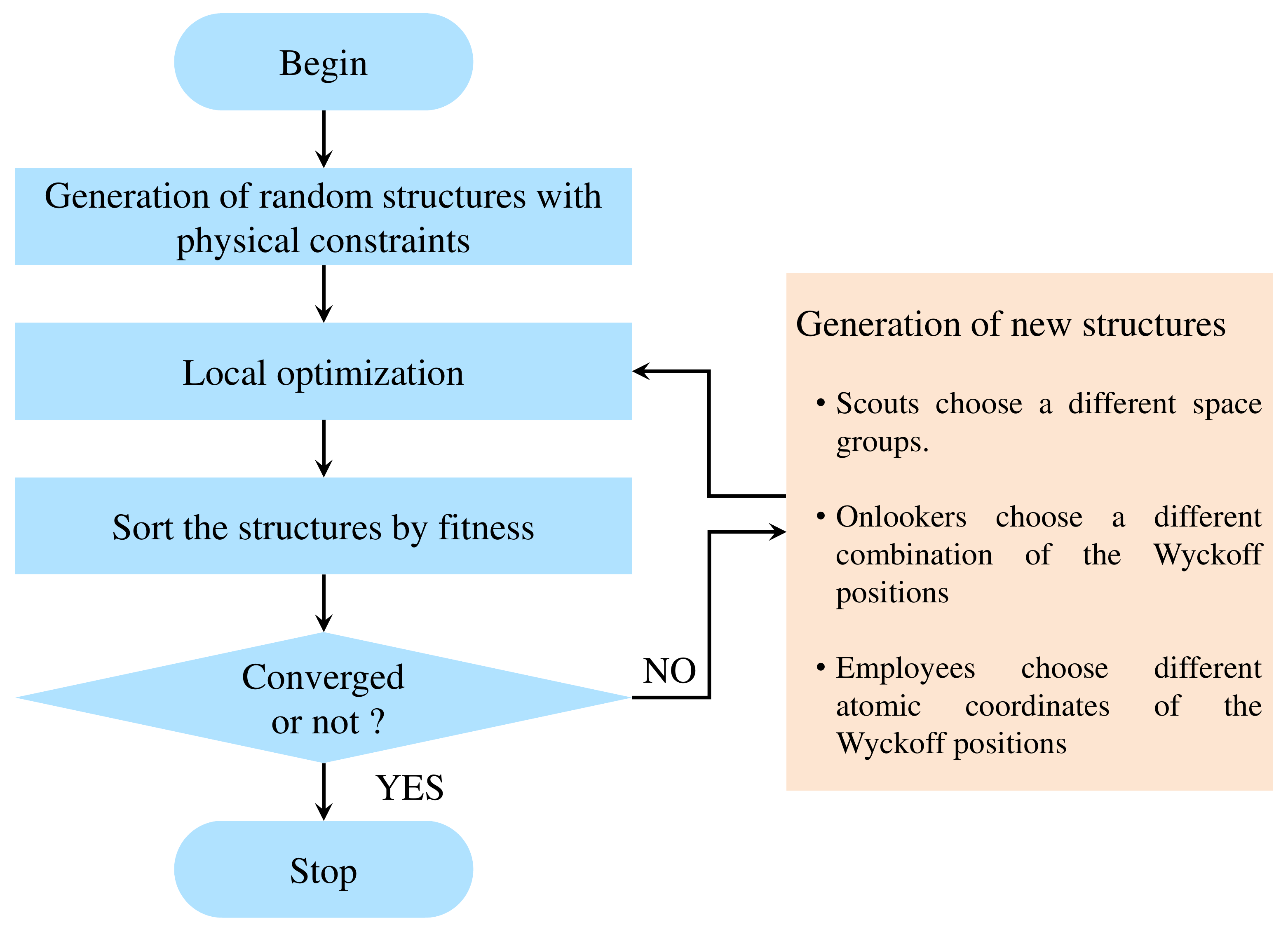}
	\caption{\label{fig:flowchart} (Color online) The flowchart of SPAI method.}
\end{figure}

Besides the general structure prediction packages, there are also some very powerful open-source programs (e.g. RandSpg\cite{avery2017randspg}, PyXtal\cite{fredericks2021pyxtal}) that can create random symmetric crystals. Our method can be easily implemented in these programs. Here, we implemented our method in the CALYPSO packages\cite{Wang_2012, wang2010crystal}. The flowchart of the SPAI method in CALYPSO is presented in Fig. \ref{fig:flowchart}. First, the initial structures are randomly generated with physical constraints that include symmetry and minimal interatomic distances. Fingerprint function\cite{Wang_2012, zhu2016fingerprint} is adopted to quantify similarities of the new structure with all the previous ones. If the structure is similar to any one of the previous, it will be discarded and replaced by a newly generated one.  After all the structures are generated, variable-cell geometry relaxations are performed to drive the structure energy to the local minimum. Then all the structures of this generation will be ranked by fitness (e.g. total energy). In the next generation, the SPAI method will be adopted to generate the new structures. These steps are iterated until a termination criterion (such as a prescribed threshold or a fixed number of iterations) is attained.
 
\section{Results and Discussion}

Our method has been benchmarked by the prediction of three known structures of MgAl$_2$O$_4$, SrTiO$_3$, and Mg$_3$Al$_2$Si$_3$O$_{12}$ having 28, 50, and 160 atoms in the lattice cells, respectively. The results are listed in Table \ref{tab:num} and compared with the results derived from the simulation runs using the previously developed local particle swarm optimization (LPSO) method\cite{Wang_2012}. Both schemes precisely reproduced the experimental structures of MgAl$_2$O$_4$ and SrTiO$_3$ with a success rate of 100\%, however, the SPAI method is more efficient than that of the LPSO method\cite{Wang_2012} as the average number of structures required to identify the true structure is much reduced.

The efficiency of the current SPAI is comparable to other popular algorithms. For example, 332 structure samplings are required to find the ground-state structure of the SrTiO$_3$ using the SPAI method, which is less than that required by other methods\cite{falls2020xtalopt,Lyakhov_2013}. Furthermore, it is evident that SPAI has excellent performance for the complex structure of Mg$_3$Al$_2$Si$_3$O$_{12}$, where the earlier approach fails without a biased input of experimental cell parameters in the simulation\cite{Lyakhov_2013}. The current method has a high success rate of 100\%, in stark contrasted to the low success rate of 20\% without any constraint of cell parameters for LPSO. It is also noteworthy that the average number of structures required to achieve the experimental structure for the current method is amazingly small at 393 for such a complex system.
\begin{table}[htbp]
		\centering
		\caption{\label{tab:num} Simulation results for MgAl$_2$O$_4$, SrTiO$_3$, and Mg$_3$Al$_2$Si$_3$O$_{12}$ derived from SPAI and LPSO algorithms without any constraint of cell parameters. 50 different structure prediction runs are performed for each system where the population per generation contains 50 structures and the maximum number of generations is set to be 100. $N$ denotes the average number of structures required to identify the experimental structures of MgAl$_2$O$_4$, SrTiO$_3$, and Mg$_3$Al$_2$Si$_3$O$_{12}$, respectively, for 50 different runs. The number of atoms in the unit cell is given in the parentheses below each system.}
		\begin{tabular*}{\columnwidth}{c@{\extracolsep{\fill}}ccc}
			\hline
			         Systems (\# atoms)        & Methods &  N   & Success rate (\%) \\ \hline
			       \multirow{2}{*}{\shortstack[c]{MgAl$_2$O$_4$\\(28)}}        &  LPSO   & 652  &        100        \\
			                                                                   &  SPAI   & 358  &        100        \\ \hline
			         \multirow{2}{*}{\shortstack[c]{SrTiO$_3$\\(50)}}          &  LPSO   & 809  &        100        \\
			                                                                   &  SPAI   & 332  &        100        \\ \hline
			\multirow{2}{*}{\shortstack[c]{Mg$_3$Al$_2$Si$_3$O$_{12}$\\(160)}} &  LPSO   & 1693 &        20         \\
			                                                                   &  SPAI   & 393  &        100        \\ \hline
		\end{tabular*}
\end{table}
 
 To demonstrate the capability of our method for applications of complex systems, we applied it to predict the plausible crystalline structures of Ba$_{1.6}$Ca$_{2.3}$Y$_{1.1}$Fe$_{5}$O$_{13}$\cite{Demont_2010}, which have been synthesized by experiments\cite{Dyer_2013}. Our approach successfully reproduced the plausible ordered structure of Ba$_2$Ca$_2$YFe$_5$O$_{13}$ containing 92 atoms in simulated cells proposed by experiments\cite{Dyer_2013} without the requirements of prior experimental knowledge, validating the effectiveness of our approach for applications to compositionally complex materials. 

Due to highly frustrated PES, identifications of the ground state crystalline structures of binary Lennard-Jones mixtures (BLJMs) pose a great challenge\cite{Middleton_2001, Amsler_2010}. Our approach is performed to determine the global minima for BLJMs containing 60, 80, and 256 atoms. The predicted structures are energetically more favourable than those found by minima hopping (MH)\cite{Amsler_2010} and basin-hopping (BH)\cite{Middleton_2001} as illustrated in Table \ref{tab:ene}. These structures share similar layered structural features (Fig. S3), which consist of simple close-packed layers formed by purely of A atoms and unexpectedly complex polyhedral layers formed by mixtures of A and B. It is notable that the unit cell dimension of the predicted structure of BLJM-256 is amazingly large over 97 angstroms. These results demonstrate that our approach holds a promise for applications to the complex structures of large systems containing more than 100 atoms.

\begin{table}[htbp]
	\centering
	\caption{\label{tab:ene} The energies of BLJM-60, BLJM-80, and BLJM-256 structures predicted by SPAI, MH, and BH. $\epsilon_{AA}$ is the potential well depth of the type A atom.}
	\begin{tabular*}{\columnwidth}{c@{\extracolsep{\fill}}ccc}
		\hline
		\multirow{2}{*}{BLJM} &    \multicolumn{3}{c}{Energy ($\epsilon_{AA}$/atom)}    \\ \cline{2-4}
		                      & SPAI  & MH\cite{Amsler_2010} & BH\cite{Middleton_2001} \\ \hline
		         60           & -7.50 &        -7.49         &          -7.08          \\
		         80           & -7.52 &        -7.50         &          -7.33          \\
		         256          & -7.47 &        -7.43         &          -7.20          \\ \hline
	 \end{tabular*}
\end{table}

\begin{figure*}[htbp]
	\centering
	\includegraphics[width=0.8\linewidth]{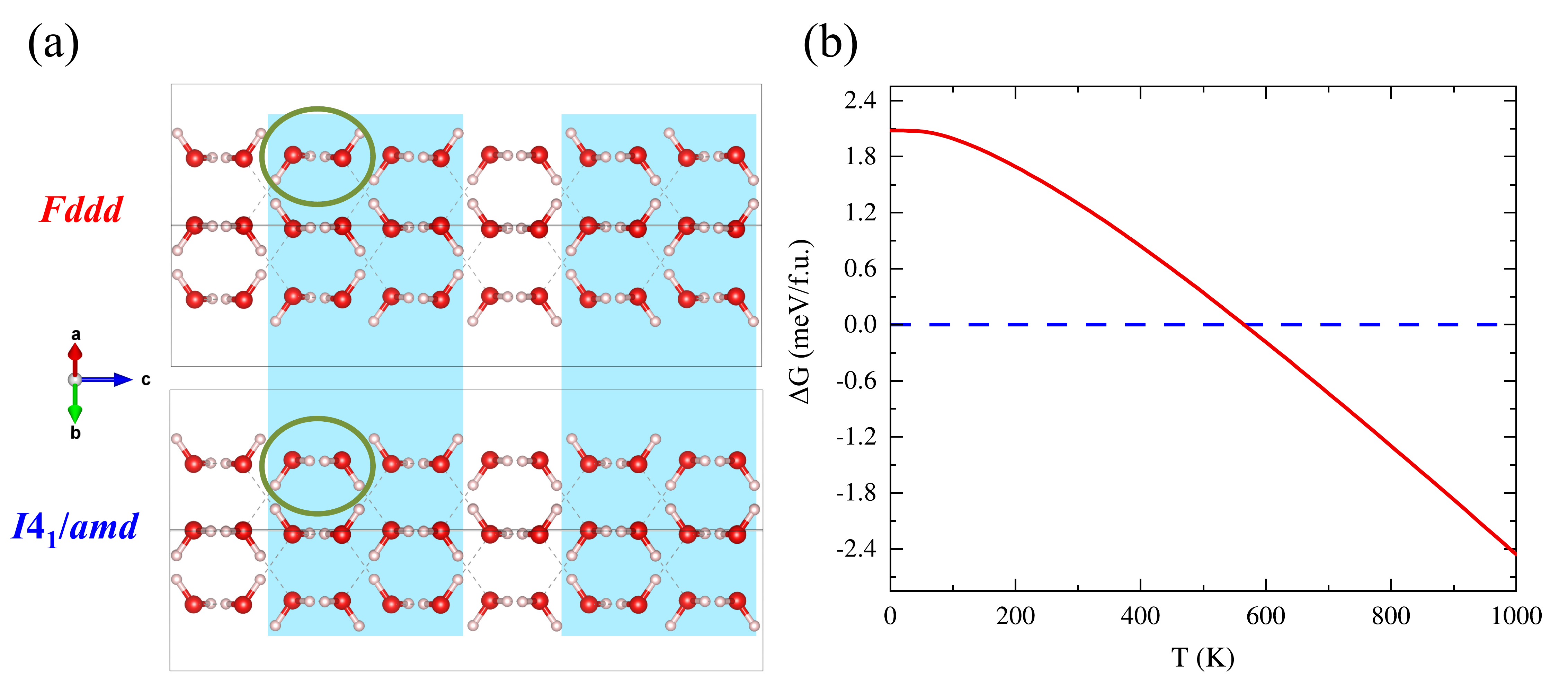}
	\caption{\label{fig:ice} (Color online) (a) The new predicted structure with a space group of $Fddd$ containing 144 atoms in comparison with the stable $I4_1amd$ structure with $\sqrt2\times\sqrt2\times3$ supercell. (b) The calculated the difference of Gibbs free energies of $Fddd$ with respective to $I4_1amd$ at a harmonic approximation level as function of temperature up to 1,000 K.}
\end{figure*}

It is expected that the crystalline structure of ice displays enormous complexity with a large unit cell because of the existence of the complex behaviour of hydrogen order/disorder. Two structure predictions of ice with simulation cells containing 48 and 144 atoms per unit cell were performed at 10 GPa using the new method and LPSO method as implemented in the CALYPSO code\cite{Wang_2012}.  The developed method successfully reproduces the experimentally observed $I4_1amd$ structure and has higher efficiency as evidenced by the fact that 208 optimized structures are required to identify the experimental structure of ice\cite{umemoto2005theoretical}, which is less than that of LPSO ($>$750 structures).  Furthermore, the developed method discoveries a new structure of $Fddd$ with the distinctive orientations of H$_2$O molecules compared with known $I4_1amd$ structure (Fig. \ref{fig:ice} (a)). The new structure contains 144 atoms per unit cell. The static energy of $Fddd$, calculated using DFT within the Perdew–Burke–Ernzerhof functional at 0 K, is higher than that of the $I4_1amd$ by only 2.0 meV/atom. The Gibbs free energies of the $Fddd$ and $I4_1amd$ are calculated within a quasi-harmonic approximation with respect to temperature up to 1,000 K. It clearly shows the stabilization of the $Fddd$ structure is energetically more favorable than $I4_1amd$ at a temperature of 600 K (Fig. \ref{fig:ice} (b)). 

Although this work is focused on the development of structure prediction on three-dimensional (3D) crystals, the proposed STG is also expected to be equally efficient for the prediction of other structures (e.g., zero-, one-, and two-dimensional structures, etc). Since there are only 17 planar or 80 layer groups, and 75 rod groups for 2D and 1D structures, respectively, structure searching simulation might be much easier. For a 0D isolated structure, point groups are used for the symmetry description, but have an infinite number. Here, the use of certain point groups $C_2$, $C_{\rm s}$, and $C_{\rm 2v}$, is expected to be useful. 

\section{Conclusion}
In summary, we have developed a CSP method by introduction of the hierarchical symmetry tree graph combined with a biased symmetry-adapted artificial intelligence algorithm. The approach drives the structure search toward the global minimum by fast identification of the most promising subsets and further in-depth exploitation. The performance of the proposed method has been demonstrated by applications to the structural complex systems of BLJMs and ice, which contain hundreds of atoms per simulated cell. As available computational resources are increased in the future, it would be expected that our method can be widely applied in the theoretical treatment of compositionally and structurally complex structures with large unit sizes containing thousands of atoms.

\section*{supplementary material}
See supplementary material for the details of the method and calculations, and information of predicted structures.

\section*{author's contribution}
X.S and J.L contributed equally to this work.

\begin{acknowledgments}
This research was supported by the National Key Research and Development Program of China under Grant No. 2016YFB0201201; the National Natural Science Foundation of China under Grants No. 11404128, 11822404, 11534003 and 11974134; Jilin Province Outstanding Young Talents project (No. 20190103040JH); Program for JLU Science and Technology Innovative Research Team; and the Science Challenge Project, No. TZ2016001. Part of the calculation was performed in the high-performance computing center of Jilin University.
\end{acknowledgments}

\section*{Author Declarations}
The authors have no conflicts to disclose.

\section*{Data Availability Statement}
The data that supports the findings of this study are available within the article and its supplementary material.

\section*{References}
\bibliography{paper}

\end{document}